\documentclass[preprint,journal]{vgtc}       

\ifpdf
  \pdfoutput=1\relax                   
  \usepackage{graphicx}                
  \DeclareGraphicsExtensions{.pdf,.png,.jpg,.jpeg} 
\else
  \usepackage{graphicx}                
  \DeclareGraphicsExtensions{.eps}     
\fi%

\graphicspath{{figures/}{pictures/}{images/}{./}} 

\usepackage{microtype}                 
\PassOptionsToPackage{warn}{textcomp}  
\usepackage{textcomp}                  
\usepackage{mathptmx}                  
\usepackage{times}                     
\usepackage{cite}                      
\usepackage{tabu}                      
\usepackage{booktabs}                  

\usepackage{fontawesome5}
\usepackage[linesnumbered,ruled,vlined]{algorithm2e}
\usepackage{algorithmicx}
\usepackage{algpseudocode}
\usepackage{amsmath}
\usepackage{tcolorbox}
\usepackage{enumitem}

\newcommand{\medley}{\textsc{Medley}}

\newcommand{\papertitle}{\medley: Intent-based Recommendations\\to Support Dashboard Composition}

\newcommand{\added}[1]{\textcolor{black}{#1}}
\newcommand{\removed}[1]{\textcolor{brown}{}}

\newcommand{\attr}[1]{\textit{#1}}
\newcommand{\intent}[1]{\textit{#1}}

\usepackage{tabularx}
\usepackage{tabu}                      
\newcolumntype{P}[1]{>{\arraybackslash}p{#1}}

\definecolor{profitgreen}{HTML}{8B8B64}

\newtcbox{\primAttrBox}{nobeforeafter, colback=gray!003, colframe=gray!25, boxrule=0.5pt, arc=1pt, boxsep=0pt,left=2pt,right=2pt,top=1.75pt,bottom=1.5pt,tcbox raise base}
\newcommand{\primAttr}[1]{\primAttrBox{{\textit{#1}}}}
\teaser{
  \centering
  \includegraphics[width=\linewidth]{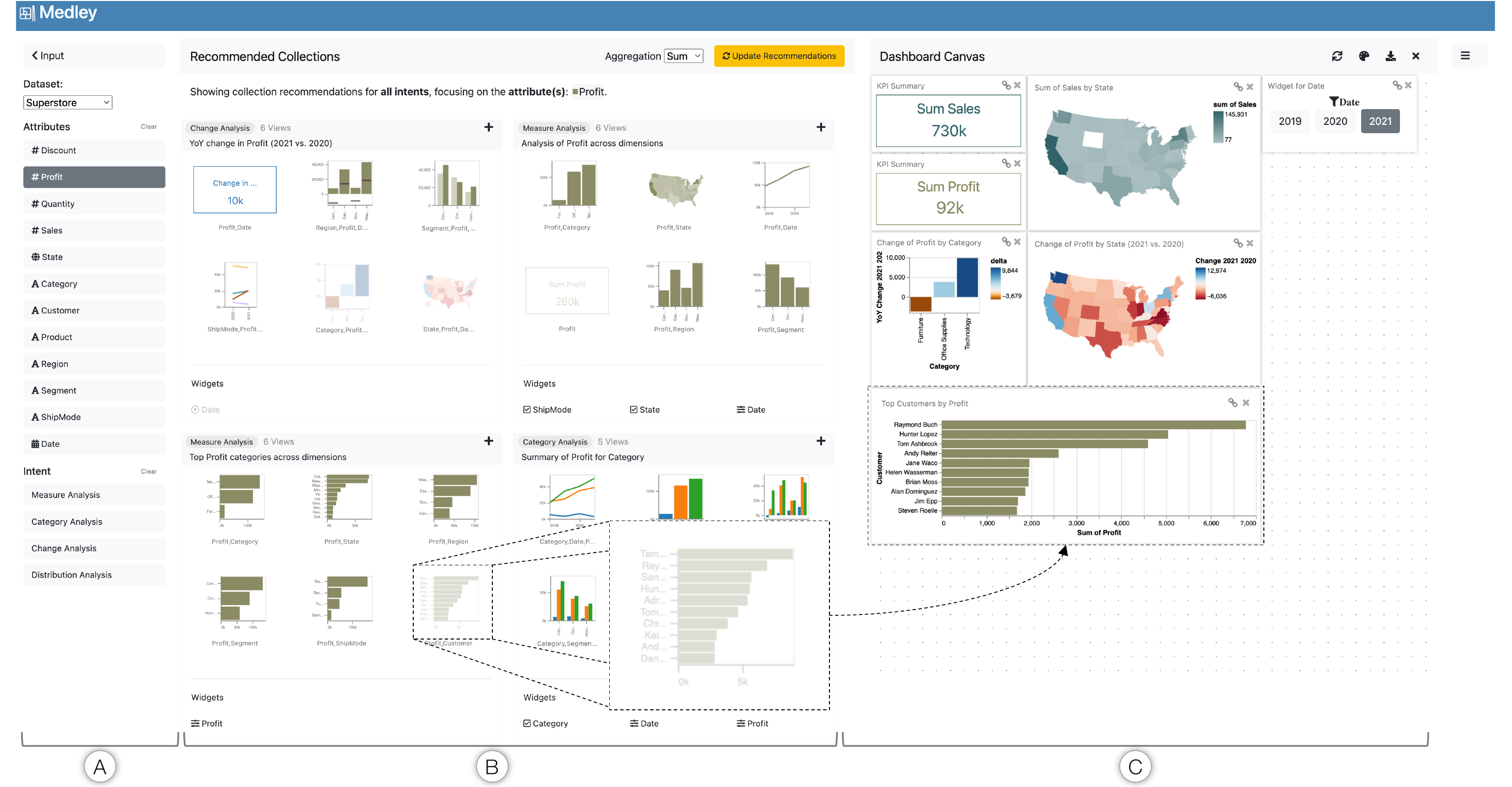}
  \vspace{-2em}
  \caption{\medley's user interface. (A) Data attribute and intent input panel, (B) collection recommendation zone, and (C) dashboard canvas. Views and widgets added to the canvas are faded out in the recommendation zone.}
  \label{fig:interface}
}

\newcommand{\figArchitecture}{
    \begin{figure}[t!]
        \centering
        \includegraphics[width=\linewidth]{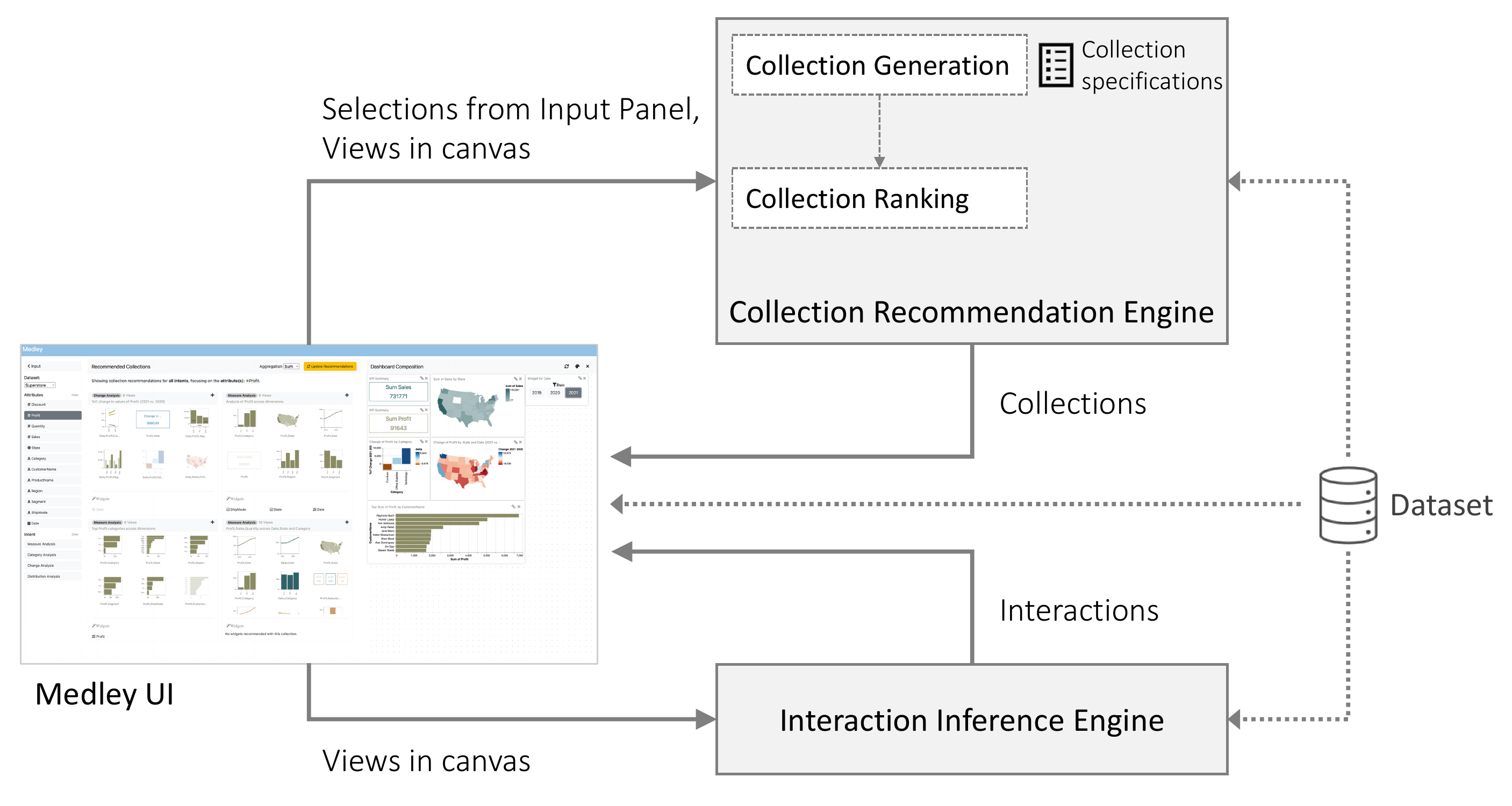}
        \vspace{-2em}
        \caption{
        System architecture overview
        }
        \label{fig:architecture}
        \vspace{-1.5em}
    \end{figure}
}

\newcommand{\figRecommendationExample}{
    \begin{figure*}[t!]
        \centering
        \includegraphics[width=\linewidth]{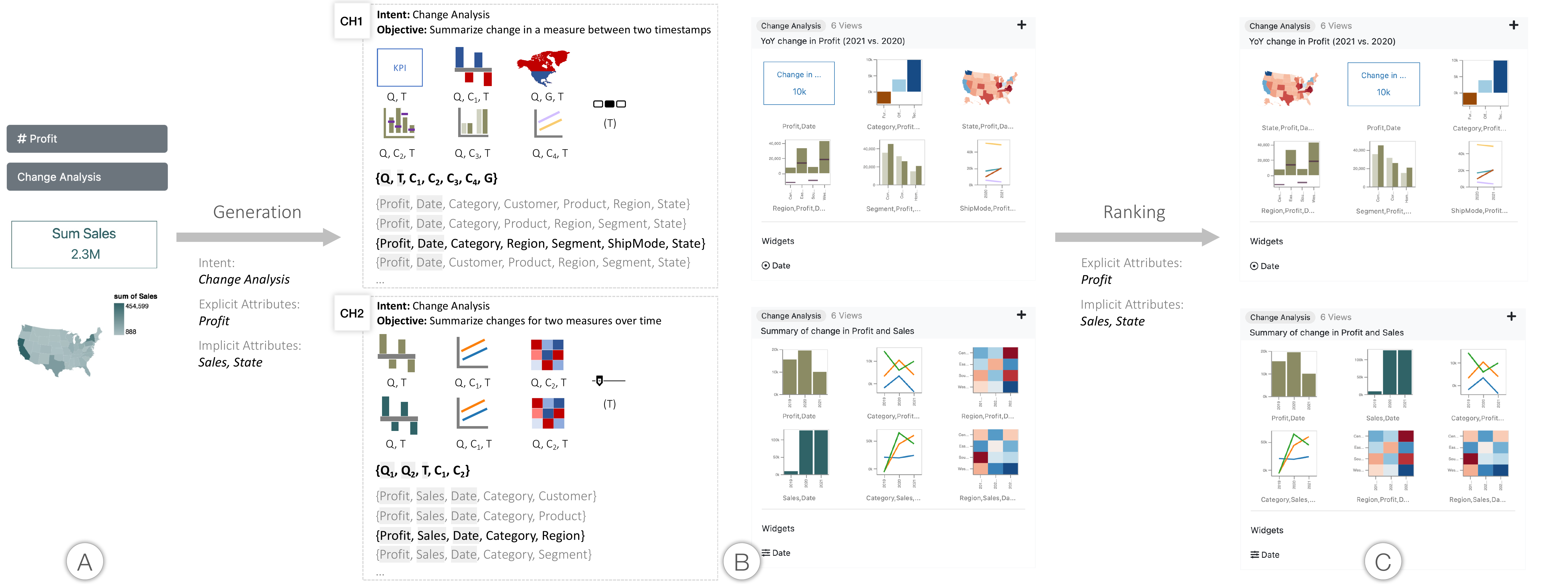}
        \vspace{-1em}
        \caption{An example illustrating \medley's recommendation process to generate recommendations shown in
        \added{Figure~\ref{fig:scenario-tall-1}C.}
        (A) Input to the recommendation engine,
        (B) collection generation to choose attribute sets and populate the change collections \texttt{CH1}, \texttt{CH2}, and
        (C) the final ranked collections.
        }
        \label{fig:recommendation-example}
        \vspace{-1em}
    \end{figure*}
}

\newcommand{\figSurveyExample}{
    \begin{figure}[t!]
        \centering
        \includegraphics[width=\linewidth]{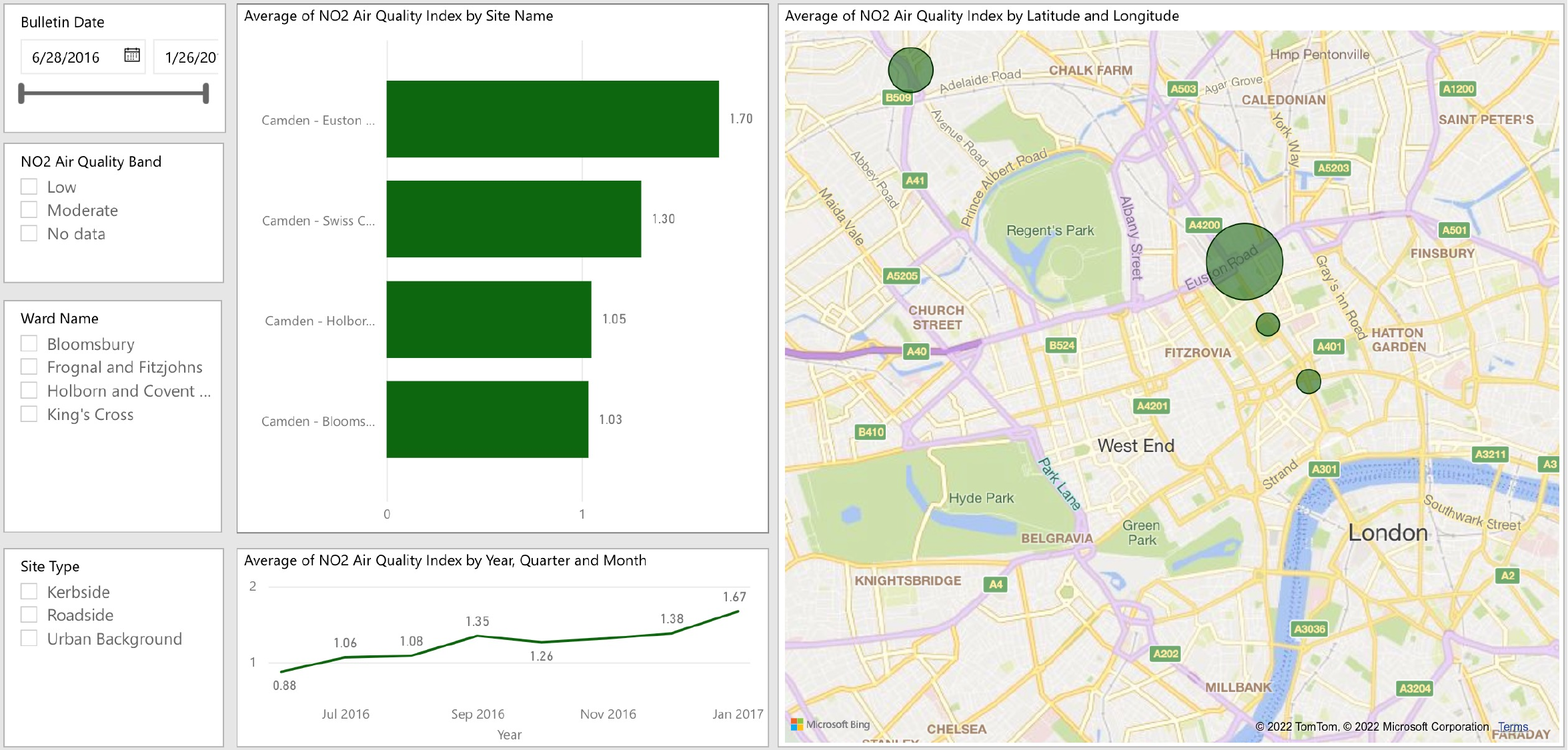}
        \vspace{-2em}
        \caption{An example dashboard~\cite{powebiDashboard2022} from our survey.
        This dashboard was mapped to the objective: ``\textit{Summarize a single measure}'' (\texttt{M1} in Table~\ref{tab:collection-specifications}) since all three views display the \attr{NO2 Air Quality Index} measure field.}
        \label{fig:survey-example}
        \vspace{-1.5em}
    \end{figure}
}

\newcommand{\figInteraction}{
    \begin{figure}[t!]
        \centering
        \includegraphics[width=.75\linewidth]{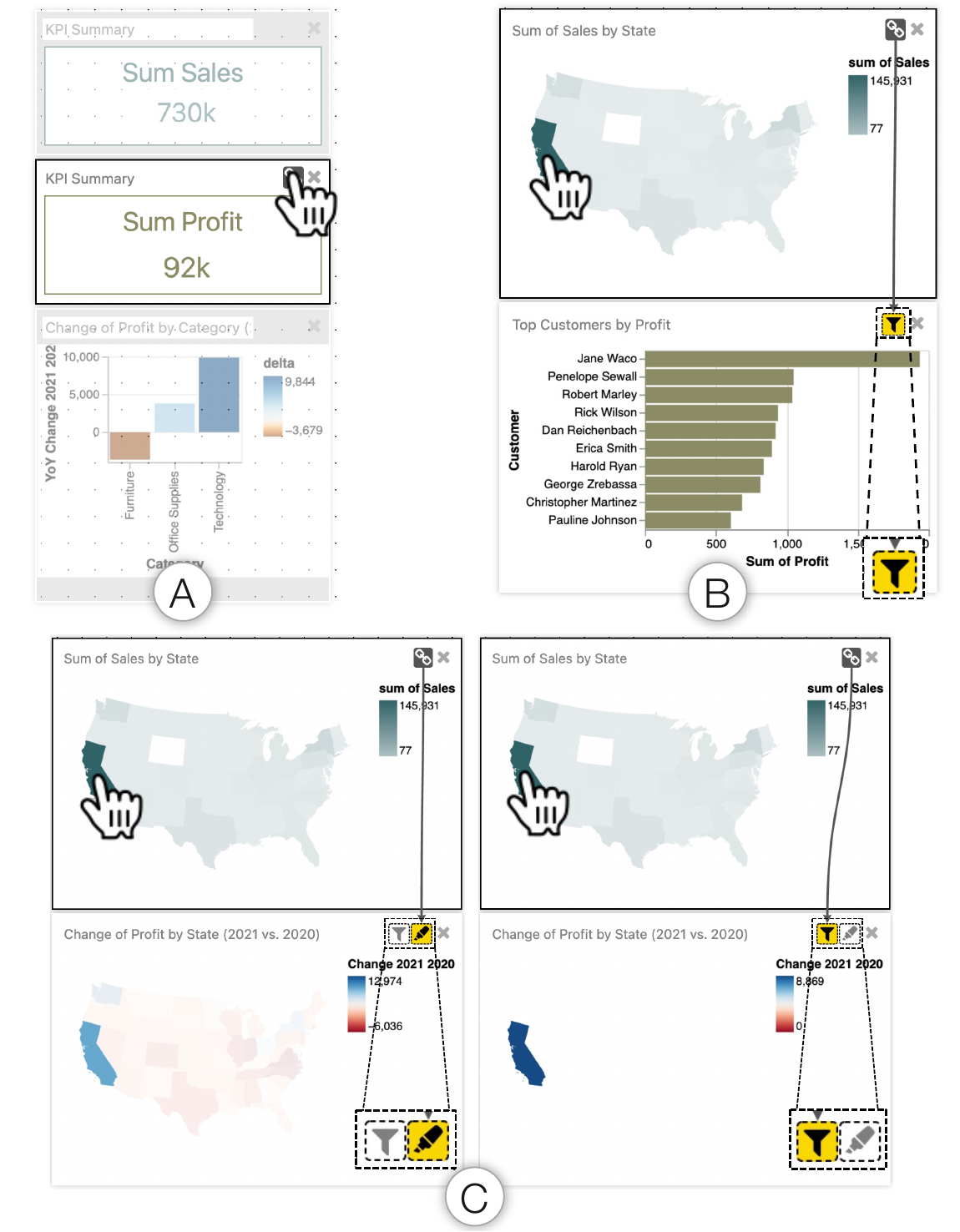}
        \vspace{-1.25em}
        \caption{\medley's dashboard interaction scenarios based on Figure~\ref{fig:interface}C.
        (A) An example where a view cannot be used to interactively update other views in the dashboard.
        (B) An example where a source view can only be used to {\small{\faFilter}}~filter items in the target view.
        (C) An example where both {\small{\faHighlighter}}~highlight and {\small{\faFilter}}~filter interactions are possible between two views.
        }
        \label{fig:interactions}
        \vspace{-1.5em}
    \end{figure}
}

\newcommand{\figTallOne}{
    \begin{figure}[t!]
        \centering
        \includegraphics[height=.91\textheight]{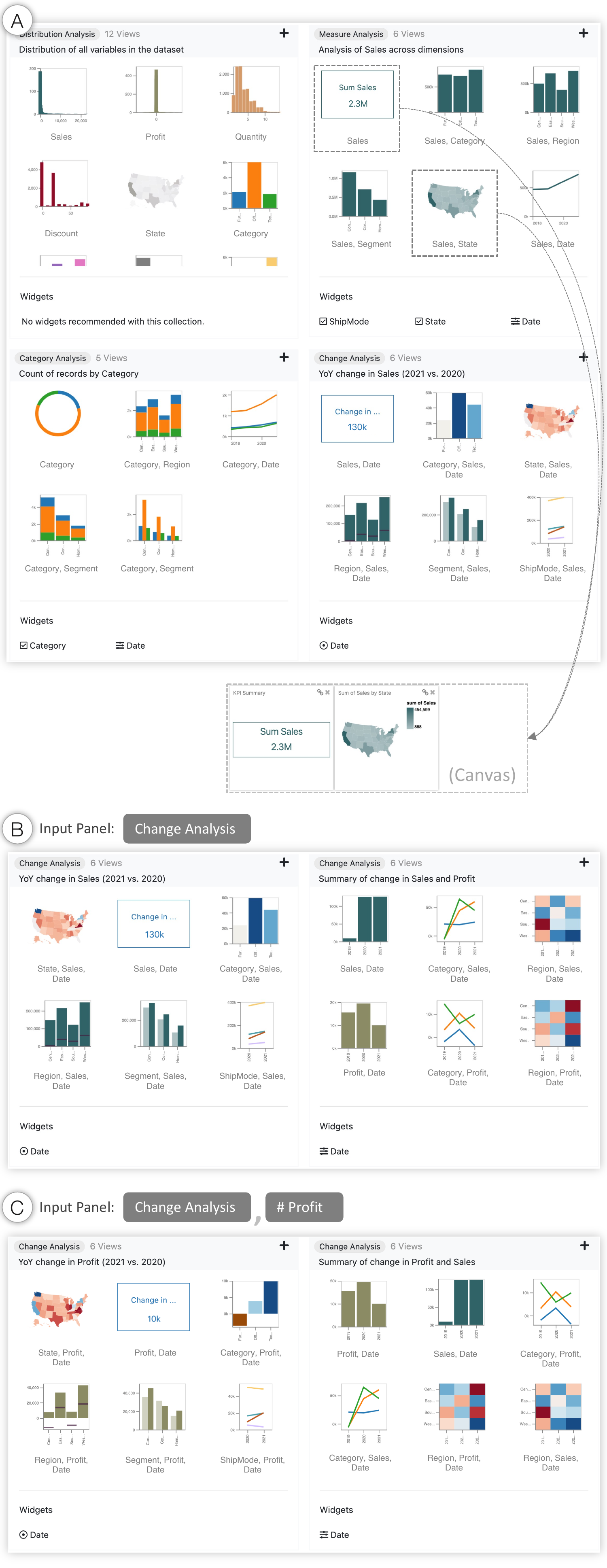}
        \vspace{-1em}
        \caption{\added{Scenes from the usage scenario. (A) Top four recommendations (one per intent) presented during a cold start. (B) Recommendations presented when the \intent{Change Analysis} intent is selected in the input panel with two views displaying \{\attr{Sales}\} and \{\attr{Sales}, \attr{State}\} already in the canvas.
        (C) Updated recommendations when the \attr{Profit} attribute is selected in the input panel.}
        }
        \label{fig:scenario-tall-1}
        \vspace{-2em}
    \end{figure}
}

\newcommand{\figStudyResults}{
    \begin{figure*}[t!]
        \centering
        \includegraphics[width=.975\textwidth]{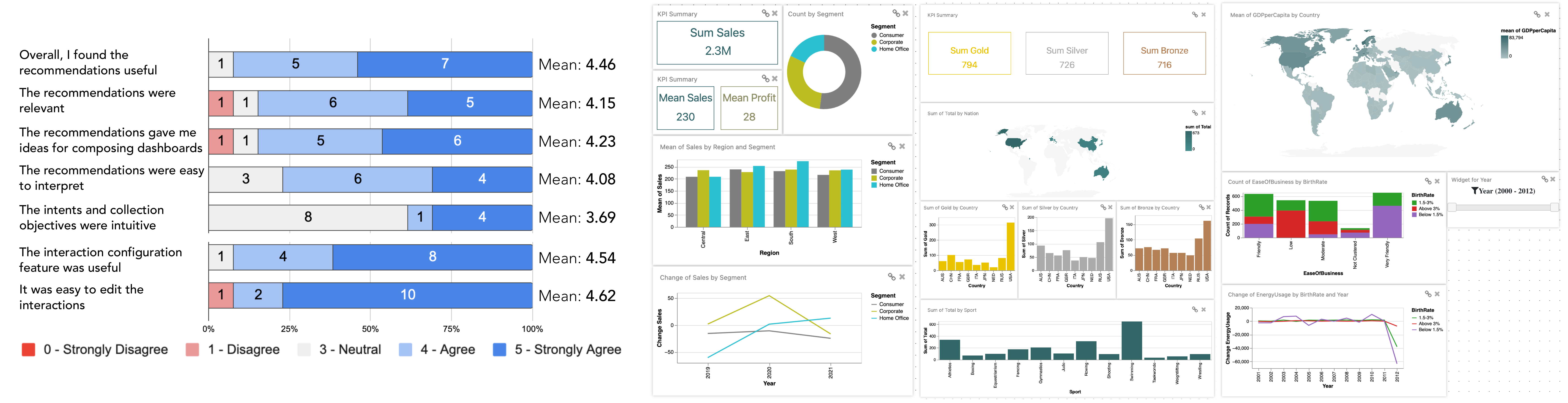}
        \vspace{-1.25em}
        \caption{Participant responses to the post-study questionnaire (left) and a sampling of dashboards composed by participants (right). The first dashboard was composed during the targeted task phase (for the prompt of creating a dashboard to summarize differences between product segments) and the other two were created during the open-ended task phase using the Olympics and the World Indicators dataset, respectively. 
        }
        \label{fig:study-results}
        \vspace{-1.2em}
    \end{figure*}
}

\onlineid{0}

\vgtccategory{Research}
\vgtcpapertype{please specify}

\title{\papertitle}

\author{Aditeya Pandey, Arjun Srinivasan, and Vidya Setlur (\textit{Member})}
\authorfooter{
\item
Aditeya Pandey is with Northeastern University. E-mail:  pandey.ad@northeastern.edu.
\item
Arjun Srinivasan and Vidya Setlur are with Tableau Research. E-mail: \{arjunsrinivasan, vsetlur\}@tableau.com.
}

\shortauthortitle{Author 1 \MakeLowercase{\textit{et al.}}: \papertitle}

\abstract{
Despite the ever-growing popularity of dashboards across a wide range of domains, their authoring still remains a tedious and complex process.
Current tools offer considerable support for creating individual visualizations but provide limited support for discovering groups of visualizations that can be collectively useful for composing analytic dashboards.
To address this problem, we present \medley, a mixed-initiative interface that assists in dashboard composition by recommending dashboard collections (i.e., a logically grouped set of views and filtering widgets) that map to specific analytical intents.
Users can specify dashboard intents (namely, measure analysis, change analysis, category analysis, or distribution analysis) explicitly through an input panel in the interface or implicitly by selecting data attributes and views of interest.
The system recommends collections based on these analytic intents, and views and widgets can be selected to compose a variety of dashboards.
\medley~also provides a lightweight direct manipulation interface to configure interactions between views in a dashboard.
Based on a study with 13 participants performing both targeted and open-ended tasks, we discuss how \medley's recommendations guide dashboard composition and facilitate different user workflows. 
Observations from the study identify potential directions for future work, including combining manual view specification with dashboard recommendations and designing natural language interfaces for dashboard authoring.
}


\CCScatlist{ 
 \CCScat{K.6.1}{Management of Computing and Information Systems}%
{Project and People Management}{Life Cycle};
 \CCScat{K.7.m}{The Computing Profession}{Miscellaneous}{Ethics}
}

\vgtcinsertpkg

\begin{document}



\firstsection{Introduction}
\maketitle

Visualization dashboards, conventionally defined as ``\textit{a visual display of the most important information needed to achieve one or more objectives}"~\cite{few2006information} are used across a range of domains including business, finance, healthcare, and sports, just to name a few. Despite their widespread use and popularity, authoring dashboards still remains a complex process. Although current visualization tools offer considerable support for creating individual views by recommending visual encodings (e.g.,~\cite{mackinlay1986automating,mackinlay2007show,wongsuphasawat2016towards,lin2020dziban}) or data fields (e.g.,~\cite{key2012vizdeck,vartak2015seedb,wongsuphasawat2015voyager,wongsuphasawat2017voyager}), they provide limited support for discovering \emph{coherent groups of views} that can be used to compose dashboards.

With limited guidance for dashboard development, users are compelled to create a series of potentially interesting individual views, identify which views comply with their dashboard's analytic objectives, and then compose a dashboard with the relevant subset of views.
Unfortunately, performing these steps can be tedious and challenging, requiring expertise in both data analysis and visualization design.
Furthermore, the sequential workflow of constructing individual views and manually composing them into a dashboard can discourage dashboard authoring and make it more of an afterthought as opposed to an active part of the users' visual analysis process.

To make dashboard composition a more active analytic process, we propose using the notion of \textbf{\emph{dashboard intents}}, or high-level analytic goals that an author wishes to facilitate through a dashboard (e.g., the dashboard should focus on summarizing data changes over time).
We posit that visualization systems can leverage intents to provide recommendations that make the dashboard authoring process more accessible to users from different domains and varying levels of data and visual analysis experience.
Specifically, we explore how intents can be used as facets to recommend \emph{coherent sets of views and filtering widgets} (hereon referred to as \textbf{\emph{collections}}) that authors can browse to select their dashboard's content from.

Through a series of interviews with dashboard authors and a survey of 200 dashboards, we identified four high-level dashboard intents---namely, \emph{measure analysis} (summarizing one or more quantitative attributes), \emph{change analysis} (showing data changes over time), \emph{category analysis} (comparing values for a categorical field), and \emph{distribution analysis} (displaying a count of records across available data fields).
We implemented \medley, a mixed-initiative interface that supports dashboard composition by recommending collections of views and widgets corresponding to different intents.
In addition to helping authors identify relevant views and widgets for their dashboards, \medley~also provides a lightweight direct manipulation-based interface for configuring interactions within the dashboard.

In this paper, we describe \medley's design and implementation, and present findings from a preliminary user study where participants used the system to perform a series of targeted- and open-ended dashboard composition tasks. 
The study results suggest merit in our premise that intent-oriented recommendations can support dashboard composition while also shedding light on the interpretability and potential user workflows with such recommendations. Finally, based on the feedback from our formative interviews and the user study, we discuss directions for future work, including complementing collection recommendations with manual view specification and designing natural language interfaces for dashboard authoring.
\section{Related Work}

\subsection{Dashboards and Multi-view Systems}

Dashboards have been a long-standing topic of interest among visualization practitioners~\cite{few2006information,wexler2017big}.
Despite their ubiquity, recent work~\cite{tory2021finding,sarikaya2018we} has highlighted that dashboards are an underexplored visualization research topic, posing a call for action to further investigate dashboards both from the author and end-user perspectives. Sarikaya et al.~\cite{sarikaya2018we} point out that a key issue for dashboard authors is the steep learning curve requiring authors to know how to operate dashboard tools as well as possess knowledge about visualization design principles.
They posit that non-expert dashboard authors, in particular, can benefit from visualization system guidance in the form of templates or recommended sets of views. Building upon this premise of guiding dashboard authors, we investigate how visualization systems can recommend dashboard collections that help authors explore relevant views and widgets to compose dashboards. Below, we discuss some existing systems that are most relevant to our work.

QualDash~\cite{elshehaly2020qualdash} generates dashboards based on templates defined to cover a set of analytic tasks specific to healthcare. GEViTRec~\cite{crisan2021gevitrec} focuses on the domain of genomic epidemiology and recommends a visually coherent combination of charts by inferring a data source graph for a set of view templates. LinPack~\cite{linkpack2022link} is a commercial product that provides hand-crafted dashboard templates for visualization tools like Tableau covering domains including retail, healthcare, and the public sector. These systems illustrate the potential of template-based recommendations to support domain-specific dashboard composition. Our proposed system also presents template-based recommendations but in a domain-agnostic setting for any tabular dataset.

Given this context, the system most related to our work is MultiVision~\cite{wu2021multivision}.
MultiVision is a deep learning-based system that recommends analytic dashboards by inferring potentially interesting data columns and choosing appropriate views. Both MultiVision and our system, \medley~are mixed-initiative interfaces that provide dashboard authoring support through view recommendations. However, there are fundamental differences between the two systems in terms of the generated recommendations. First, MultiVision's recommendations are geared toward providing an analytic overview, whereas \medley's recommendations are designed to support targeted dashboard intents (e.g., comparing measures, analyzing changes over time).
Second, while MultiVision recommends a single set of coordinated views, \medley~recommends multiple collections of coordinated views and widgets. In doing so, \medley~gives users the flexibility to compose a dashboard directly from one of its recommendations or by mixing views and widgets from different collections. \medley~also adds support to configure interactions between a dashboard's elements.

Our work also relates to prior research on the broader topic of multiple-view visualization design, a summary of which can be found in other research papers (e.g.,~\cite{al2019towards,chen2020composition,roberts2007state}).
While we do not make novel research contributions in the space of multi-view visualization design, we incorporate guidelines from prior work~\cite{qu2017keeping,wang2000guidelines,roberts1998encouraging,kristiansen2021semantic,qu2017keeping} in designing our recommendations.

\subsection{Task-driven Visualization Recommendation}

\added{There is a plethora of research on visualization recommendation techniques and systems, a detailed review of which can be found in other survey manuscripts~\cite{collins2018guidance,lee2021deconstructing,zhu2020survey,wu2021survey}.}
The key idea of our work, however, is the notion of a dashboard's analytic \emph{intent}, which is similar to the concept of visual analysis \emph{tasks} proposed in prior literature (e.g.,~\cite{amar2005low,brehmer2013multi,schulz2013design,sarikaya2017scatterplots}).
Given this overlap, we discuss examples from one subset of prior research on visualization recommendation that is most relevant to our work: task-driven recommendation systems that consider one or more analytic tasks (e.g., correlate, analyze trend) in addition to data attributes when recommending visualizations.

Casner~\cite{casner1991task} presents one of the earliest examples of visualization systems that suggests charts based on a user's task (e.g., using a circular graph layout to find direct flight routes or a table to see flight information).
Gotz and Wen~\cite{gotz2009behavior} present a prototype system that observes interaction patterns (e.g., repeatedly changing filters or swapping attributes) to infer analytic tasks such as comparison or trend analysis and correspondingly recommends visualizations such as small multiples or line charts.
VizAssist~\cite{bouali2016vizassist} enables its users to specify their data objectives in terms of analytic tasks (e.g., correlate, compare) and considers these tasks in combination with existing perceptual guidelines as input to a genetic algorithm for recommending visualizations. Saket et al.~\cite{saket2018task} conduct a study to assess the effectiveness of five canonical visualizations on ten low-level analytic tasks~\cite{amar2005low} and developed the Kopol recommendation engine based on their study's findings.
Kim and Heer~\cite{kim2018assessing} discuss how analytic tasks and data distributions, in addition to the choice of visual encodings, could determine the effectiveness of a visualization.
TaskVis~\cite{shen2021taskvis} is another recent system that consolidates findings from prior recommendation systems and perceptual studies to recommend task-based visualizations and supports four ranking schemes to drive the recommendations.

These systems all explore the idea of task-based recommendation, but for generating \textit{individual visualizations}. Extending beyond individual views, we explore how the notion of task- or intent-based recommendations can be extended to a dashboard composition context, considering the nuances of recommending coherent yet diverse collections of multiple views and widgets.
In doing so, we identify both a preliminary set of analytic intents for dashboards as well as representative collections of views and widgets that map to those intents.

\figTallOne

Findings from these task-based recommendation systems and studies have also been used to develop systems that leverage analytic tasks as facets for organizing recommendations.
For example, Foresight~\cite{demiralp2017foresight} uses tasks like distributions, outliers, and correlations to guide insight discovery, grouping recommendations involving histograms, box plots, and scatterplots, respectively.
Similar to Foresight, systems like Voder~\cite{srinivasan2018augmenting} and Datasite~\cite{cui2019datasite} also recommend visualizations but encompass them within textual data facts/insights (e.g., ``\textit{Displacement and Horsepower have a strong correlation}'') that, in turn, correspond to analytic tasks (e.g., \textit{correlation}).
Frontier~\cite{lee2021deconstructing} is another example of a system that categorizes its recommended visualizations within different tasks (e.g., filter, distribution, correlation) to guide data exploration.
User studies of these systems have shown that using tasks or intents as an organizing principle facilitates various data exploration strategies while also aiding interpretation of the recommendations themselves.
Drawing inspiration from these systems, we use intents as a way to both generate and organize \medley's recommendation and overcome potential interpretability challenges that accompany prior multi-view visualization recommendation systems~\cite{wu2021multivision}.
\section{Usage Scenario}
\label{sec:usage-scenario}


We first describe a usage scenario to motivate \medley's design and illustrate how the tool enables dashboard composition. The scenario uses a product sales dataset\footnote{
The scenario video along with the Superstore dataset are included as part of the supplementary material.}
for a fictitious company, ``Superstore.''

Each row in the dataset is a sale and there are 12 columns (shown in Figure~\ref{fig:interface}A) including four quantitative attributes (e.g., {\small{\faHashtag}}~\attr{Sales}, {\small{\faHashtag}}~\attr{Profit}), six categorical attributes (e.g., {\small{\faFont}}~\attr{Category}, {\small{\faFont}}~\attr{Segment}), one geographic attribute ({\small{\faGlobe}}~\attr{State}), and a temporal attribute ({\small{\faCalendar}}~\attr{Date}).
For consistency, we use this dataset as an example throughout the paper. 

Imagine Sarah, an operations-reporting analyst who designs dashboards for the Superstore company's executives.
For an upcoming annual board meeting, Sarah needs to compose a dashboard that reviews the company's performance.

\vspace{.5em}
\noindent\textbf{Starting with system recommendations.}
Sarah starts by inspecting \medley's recommended collections (Figure~\ref{fig:scenario-tall-1}A).
Glancing through the recommendations, Sarah reads the second collection's description ``\textit{Analysis of Sales across dimensions}'' and peruses its views.
To provide an overview of \attr{Sales}, she adds two views---the data summary view and the choropleth map to the dashboard canvas.
Skimming through other recommendations, Sarah is intrigued by the fourth \intent{Change Analysis} collection that displays ``\textit{Year-over-year (YoY) change in \attr{Sales} (2021 vs.~2020)}'' and considers creating a dashboard that displays an overview of the current year's values but also provides context by summarizing changes since the previous year.

\vspace{.5em}
\noindent\textbf{Narrowing in on an intent while updating the dashboard.}
To see similar collections, Sarah selects the \intent{Change Analysis} intent in the input panel.
In response, \medley~filters its recommendations to only show two collections (Figure~\ref{fig:scenario-tall-1}B).
The first collection displays views for changes in \attr{Sales} one year at a time, and the second collection displays yearly changes in \textit{Sales} and \textit{Profit} across all years in the dataset. This second collection gives Sarah the idea of using both \attr{Sales} and \attr{Profit} for her dashboard.
However, she wants to preserve the \textit{Sales} overview charts currently in her canvas and complement those with views that show the change in \attr{Profit} one year at a time. Sarah selects \attr{Profit} in the input panel and sees an updated set of recommendations 
\added{(Figure~\ref{fig:scenario-tall-1}C)}.
From the new ``\textit{YoY change for \attr{Profit} (2021 vs.~2020)}'' collection, Sarah adds the choropleth map and the difference bar chart to the canvas. She also adds the suggested year picker widget to display values and changes one year at a time in the dashboard.

\vspace{.5em}
\noindent\textbf{Providing an overview.}
Realizing that the dashboard lacks an overview of \attr{Profit}, Sarah deselects the \intent{Change Analysis} intent from the input panel to see other views and collections.
Looking at the updated set of recommendations
\added{(Figure~\ref{fig:interface}B)}, Sarah sees a data summary view of the total \attr{Profit} in the second collection and adds the view to her canvas.
Intrigued by the new ``\textit{Top \attr{Profit} categories across dimensions}'' collection (third collection in
\added{Figure~\ref{fig:interface}B}),
she inspects its views and also adds a sorted bar chart showing the 10 most profitable \attr{Customer}s.

\vspace{.5em}
\noindent\textbf{Finalizing dashboard interactions and layout.}
As Sarah adds views to the canvas, \medley~automatically configures interactions between the dashboard elements (clicking a view to filter other views, updating all views when a year is selected in the widget).
With an interactive set of six views and a widget, Sarah decides that she has a dashboard that can help answer a breadth of analytic questions (e.g., \textit{What are the total Sales for the most recent year? Which states contributed most to those Sales?
Were there states that suffered a loss even though they had high Sales?}).
She proceeds to adjust the view sizes and layout to make the dashboard more presentable and shares it with her team.
\section{Design Process and Goals}

\begin{table*}[t!]
    \centering
    \caption{
    Collection specifications currently supported in \medley.
    The letters in the attribute mapping column indicate attribute types, \textbf{Q}: quantitative, \textbf{C}: categorical, \textbf{G}: geographic, and \textbf{T}: temporal.
    \primAttr{Attributes} with a gray background are the primary attributes that are required to generate a collection.
    Other attributes listed as plain text are secondary attributes that may be skipped if they are unavailable in the dataset.
    For example, for the \texttt{M1} collection with the attribute mapping \{\primAttr{Q}, \attr{3C}, \attr{G}, \attr{T}\}, if a dataset only had two categorical fields or did not have a geographic attribute, \medley~would generate two bar charts instead of three or skip recommending a map, respectively. View icons in the fourth column are color coded to match the primary attributes they visualize or are given a diverging red-blue color if they show changes. 
    Specifically for the \texttt{CAT1} and \texttt{CAT2} collections, \attr{Q} can either be a quantitative attribute or the count of records in a dataset (where views include donut charts and stacked bar charts\added{, as shown in the third collection of Figure~\ref{fig:scenario-tall-1}A}). \added{Icons in the rightmost column represent the suggested data selection and filtering  widget types.}
    }
    \includegraphics[width=\textwidth,keepaspectratio]{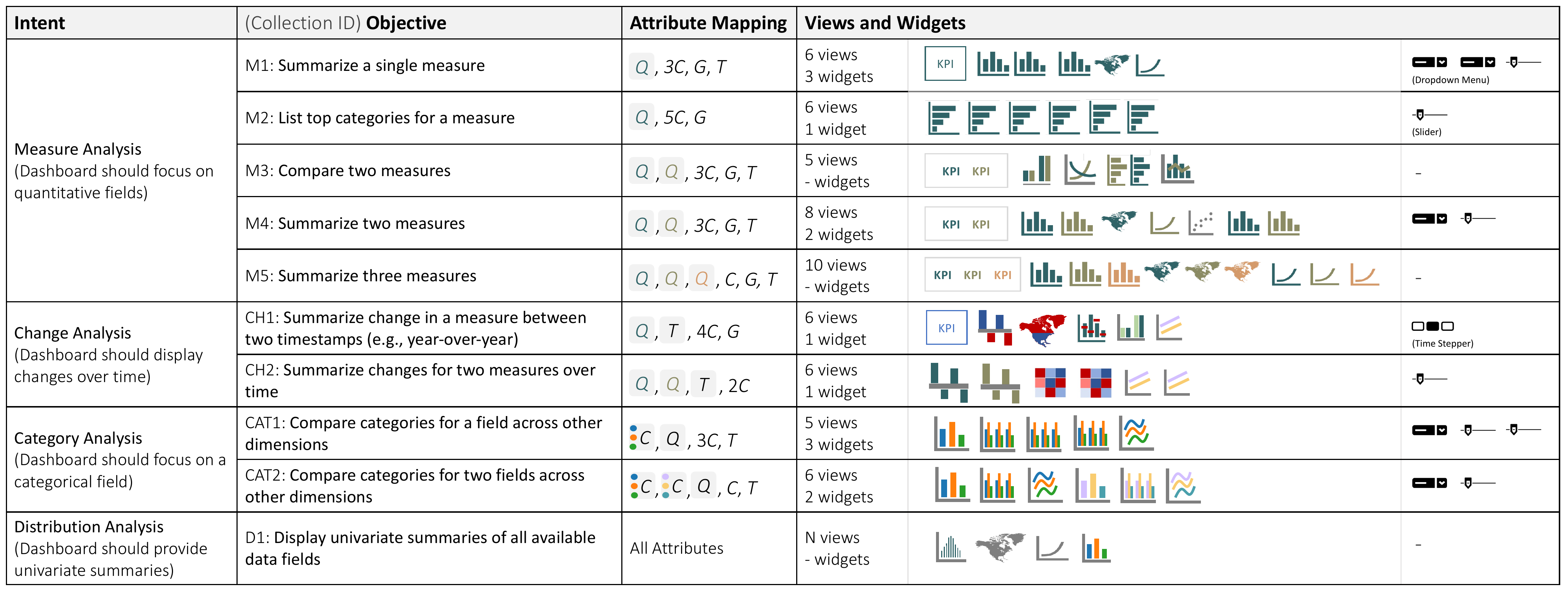}
    \label{tab:collection-specifications}
    \vspace{-2.5em}
\end{table*}

\medley's~design and user experience was informed by a set of interviews with dashboard authors and surveying a variety of dashboards.

\subsection{Identifying Dashboard Intents and Collections}

Our work builds upon the premise that dashboards are not just a miscellaneous set of views but rather a coherent group of views to address specific analytic intents.
To validate this premise, we interviewed four employees (two analysts and two consultants) at a data analytics software company to better understand dashboard authoring processes.
The interviewees were recruited through mailing lists of design and reporting teams who authored dashboards at a daily- or weekly-basis.

When asked about their experience authoring dashboards, the interviewees stated that in most situations, they would receive a set of requirements from their managers or clients (i.e., the consumers of the dashboard) in the form of specific data attributes (e.g., ``\textit{We are interested in getting an overview of the company's sales for the past year}") or analytic questions the dashboard should help answer (e.g., ``\textit{Are there product areas the company should invest in from a marketing perspective? What are the potential returns?}''). They would then use the attributes and questions to think about goals for their dashboard (e.g., the dashboard should provide an overview of sales followed by a breakdown of the sales by different product segments and geographic regions) and subsequently look for views that fit those goals.

In other situations, when interviewees encountered a dataset for the first time or did not have a clear set of requirements, they said that they perform exploratory analysis to generate a slew of visualizations and then start combining subsets of the generated views into coherent themes to create dashboards. While these two situations represent disparate workflows (the first is more targeted, whereas the second is more open-ended), the idea that the eventual dashboard should emphasize a specific goal (or intent) helped validate our core premise.

To identify a list of dashboard intents and representative collections of views and widgets that map those intents, we curated a list of 200 dashboards from Tableau~\cite{tableaupublic2021link} and Microsoft Power BI's~\cite{microsoftpbi2021link} dashboard galleries.
During the process of curating the dashboards, we covered a breadth of popular data domains, including business \& finance, health, human resources, and sports, among others.
To scope our search and initial prototype design, we only focused on dashboards that consisted of two or more basic chart types (e.g., bar chart, map, large numbers).

Two of the authors collaboratively inspected the dashboards to identify an initial list of \textit{\textbf{objectives}} that describe specific analytical tasks (e.g., summarizing a single measure field, facilitating comparison of two measure fields, displaying changes in multiple measure fields over time) based on the dashboards' title, views, widgets, interactions, and other meta-data such as textual description (e.g., Figure~\ref{fig:survey-example}).
We then iteratively refined the dashboard groups and objectives and mapped lower-level analytic objectives to broader categories of analytic intents. 

\figSurveyExample

Table~\ref{tab:collection-specifications} lists the four high-level intents and the 10 more targeted analytic dashboard objectives that we identified through the survey and iterative coding process.
Each collection (row) in Table~\ref{tab:collection-specifications} maps to a high-level intent (first column), a more focused analytic objective (second column), and displays a specific combination of attribute types (third column) using a list of views and widgets (fourth column). Notations in the second column (e.g., \texttt{M1}, \texttt{M2}, \texttt{CH1}) are used to reference collections in other figures and text in the paper.

For each objective, we also identified the most frequently occurring view types and widgets in the surveyed dashboards. Besides frequency of occurrence, we also followed Wang et al.'s guidelines~\cite{wang2000guidelines} to ensure that the resulting dashboards do not have an overwhelming number of views (e.g., collections focus on 1-3 primary attributes and contain no more than 10 views) and that the views collectively facilitate data understanding and querying (e.g., views cover different combinations of attribute and chart types and can be interactively linked).
Furthermore, we considered widgets in addition to views as our interviewees stated that interactivity is a key feature of dashboards, and filtering widgets are commonly used to facilitate analytical inquiry and exploration.

Note that the intents, objectives, and collections in Table~\ref{tab:collection-specifications} are not mutually exclusive \added{(i.e., a dashboard may combine two or more intents)} or exhaustive \added{(i.e., new intents or objectives may be discovered as more dashboards are surveyed}), and are not intended to be a prescriptive list. Rather, the survey was primarily to help ground our design and develop an initial prototype to elicit user feedback.

\subsection{Design Goals}

With an initial set of intents and collections, we followed an iterative design process to develop \medley.
Over a span of six weeks, we met multiple times with the four dashboard authors we interviewed earlier and recruited three additional participants (two designers and a product manager for a dashboard authoring tool) through the same mailing list.
The three new participants were less versed in dashboard authoring but were knowledgeable about the requirements of dashboard tool users with different levels of visualization expertise.
During the meetings, we briefly introduced the interviewees to the idea of dashboard intents and objectives, and demonstrated working versions of \medley's prototype to gather feedback.
Combining the feedback from these meetings with guidelines and findings from prior work on visualization recommendation systems (e.g.,~\cite{key2012vizdeck,wongsuphasawat2015voyager,crisan2021gevitrec,wu2021multivision}), mixed-initiative interfaces~\cite{horvitz1999principles,hearst1999mixed}, and multi-view visualization design (e.g.,~\cite{qu2017keeping,wang2000guidelines}), we iteratively compiled a list of seven goals that guided \medley's final design and implementation.

\vspace{.5em}
\noindent\textbf{DG1. Support flexibility in dashboard composition.}
During the design sessions, all interviewees found the collection themes and recommendations to be intuitive and helpful.
However, the four expert authors noted that dashboard composition is often subjective, and the views in a single collection may not all map to what the author has in mind for the dashboard. To enable authors to choose the recommendations in flexible ways, \medley~allows for adding entire collections or selecting individual views and widgets across different collections.

\vspace{.5em}
\noindent\textbf{DG2. Support explicit user specification of attributes and intents.}
Our interviewees noted that dashboard authors are often interested in specific data attributes or intents, and this interest can vary over the course of a session. Thus, similar to prior visualization recommendation systems (e.g.,~\cite{wongsuphasawat2015voyager,key2012vizdeck,bouali2016vizassist,wu2021multivision}), \medley~allows users to provide their preferences by explicitly selecting attributes and intents in the input panel, and internally uses these selections to drive its recommendations.

\vspace{.5em}
\noindent\textbf{DG3. Support context-driven recommendations.}
In an initial version of \medley, only attribute selections in the input panel were used for generating recommendations.
However, as participants added content to the canvas, they found the recommendations to be less useful over time and expressed the need for the system to also consider the context of the views already in the canvas. Hence, besides \textit{explicit} user selections in the input panel (\textbf{DG2}), \medley~also infers \textit{implicit} user input from views added to the canvas while generating recommendations.

\vspace{.5em}
\noindent\textbf{DG4. Select default attributes based on data interestingness.}
Participants noted that when they do not have specific attributes or intents in mind, they would benefit from system recommendations as a starting point or to explore alternative dashboard compositions.
To recommend views and collections that display potentially interesting data patterns, \medley~leverages statistical metrics from prior insight-based visualization recommendation systems (e.g.,~\cite{cui2019datasite,demiralp2017foresight,srinivasan2018augmenting}).

\vspace{.5em}
\noindent\textbf{DG5. Apply multi-view design guidelines in recommendations.}
During the interviews, the designers and product manager stated that dashboard tool users, particularly novices, often lack knowledge about good dashboard design practices.
To promote consistency, \medley's~recommendations incorporate multi-view design guidelines~\cite{qu2017keeping,wang2000guidelines} by default such that all views have a zero baseline, a consistent sort order within each collection, and consistent color encoding used for attributes across collections.

\vspace{.5em}
\noindent\textbf{DG6. Provide descriptive text to aid interpretation.}
While the interviewees found grouping views and widgets into collections to be useful for visual scanning, they also expressed the need for textual descriptions to summarize the content within individual collections for better understanding.
Interviewees also suggested using terms like ``measure'' (a common BI term used to refer to quantitative attributes) and ``dimension'' (a common BI term used to refer to categorical and temporal attributes) so that the text is more familiar with users of mainstream dashboard authoring tools.

\vspace{.5em}
\noindent\textbf{DG7. Support for interactions during dashboard composition.}
Interactivity can enable dashboard viewers to explore and answer a broader spectrum of data questions. Similar to prior studies (e.g.,~\cite{wu2021multivision}), our interviewees found understanding the connectivity between dashboard elements and choosing an interaction technique (e.g., highlight vs.~filter) to be a challenge with existing tools. Addressing this need, \medley~allows for the inspection and direct, graphical editing of interactions between dashboard elements in the interface.
\section{Medley}

\figArchitecture

\figRecommendationExample

Figure~\ref{fig:interface} shows an overview of \medley's interface.
The three main components are: (A) an input panel, (B) a list of collection recommendations, and (C) the dashboard canvas.
The recommended collections are collapsible
and are sorted by relevance from top to bottom and left to right (users can vertically scroll down to see lower-ranked collections).
Short text descriptions summarize the objective of individual collections (e.g., ``YoY Change in Sales (2021 vs. 2020)" in Figure~\ref{fig:scenario-tall-1}B) (\textbf{DG6}).
\added{Views and widgets within each collection are shown as thumbnails to optimize for space and to facilitate rapid browsing. Hovering on thumbnails displays tooltip previews of the views and widgets.}
\added{To promote visual consistency in attributes across views and collections (\textbf{DG5}), the system assigns a unique color to each quantitative attribute (e.g., \texttt{M5} in Table~\ref{tab:collection-specifications}), a unique color to each value in a categorical attribute (e.g., \texttt{CAT1} in Table~\ref{tab:collection-specifications}), and uses a diverging red-blue color scale for views that directly encode change (e.g., \texttt{CH1} in Table~\ref{tab:collection-specifications}).}
\added{Users can add a view or widget to the canvas by double-clicking on a thumbnail or add an entire collection by clicking the {\small{\faPlus}} icon in the collection header (\textbf{DG1}). Items on the canvas can be dragged and resized.}

\medley~is implemented as a web-based application developed using HTML/CSS and JavaScript.
The system accepts as input tabular datasets containing {\small{\faHashtag}} quantitative, {\small{\faFont}} categorical, {\small{\faGlobe}} geographic, or {\small{\faCalendar}} temporal attributes. Visualizations in \medley~are created using Vega-Lite~\cite{satyanarayan2016vega}.
Figure~\ref{fig:architecture} presents a high-level overview of the system's architecture.
We now describe the Collection Recommendation Engine and the Interaction Inference Engine components in more detail.

\subsection{Collection Recommendation Engine}

\medley~recommends collections either at the start of a session (e.g., Figure~\ref{fig:scenario-tall-1}A), when attributes or intents are selected in the input panel (e.g., Figure~\ref{fig:scenario-tall-1}B,C), or when users click the \textit{{\small{\faSync}}~Update Recommendations} button (top-right corner of Figure~\ref{fig:interface}B) after adding views and widgets to the canvas. Algorithm~\ref{algo:recommendation} provides an overview of \medley's~recommendation logic.

To generate recommendations most relevant to the user's interest and interaction (\textbf{DG2}, \textbf{DG3}), the system first determines the set of attributes the user is interested in by inspecting the input panel for \textit{explicit} attribute selections as well as views in the canvas for \textit{implicit} attribute references (Algorithm~\ref{algo:recommendation}, lines 4-7).
For brevity, hereon, we refer to the combined set of explicit and implicit attributes as ``\emph{attributes of interest}.'' Using the attributes of interest and the canvas views, \medley~follows a two-step generation and ranking process to provide recommendations (illustrated in Figure~\ref{fig:recommendation-example}).

\begin{algorithm}[t!]
    collections = []\\
    dataAttrs = [dataset attributes]\\
    collectionSpecifications = [M1, M2,..., D1 in Table~\ref{tab:collection-specifications}]\\\vspace{.5em}
    canvasViews = [views in dashboard canvas]\\
    explicitAttrs = [attributes selected in the input panel]\\
    implicitAttrs = getAttributes(canvasViews)\\
    attrsOfInterest = [explicitAttrs + implicitAttrs]\\
    otherAttrs = [dataAttrs - attrsOfInterest]\\\vspace{.5em}
	userIntents = [intents selected in input panel] // defaults to all four intents\\
	\For{c in collectionSpecifications}{
		\uIf{c.intent in userIntents}{
			populateCollection(c, attrsOfInterest, otherAttrs)\\
			collections.push(c)\\
		}
	}\vspace{.5em}
	rankCollections(collections, attrsOfInterest, canvasViews)\\
	\textbf{return} collections
	\caption{recommendCollections()}
    \label{algo:recommendation}
\end{algorithm}

\subsubsection{Collection Generation}

During collection generation (Algorithm~\ref{algo:recommendation}, lines 9-13), \medley~enumerates a list of collections by populating the collection specifications listed in Table~\ref{tab:collection-specifications}. The system first checks if a user has selected any intents in the input panel. If no intents are selected, \medley~populates collections for all four intents (\intent{Measure Analysis}, \intent{Change Analysis}, \intent{Category Analysis}, and \intent{Distribution Analysis}). If one or more intents are selected, the system only considers collection specifications corresponding to those intents (\textbf{DG2}). \medley~then populates individual collections by iterating through a list of possible attribute sets based on the attribute mappings in Table~\ref{tab:collection-specifications}. 

Consider the example in Figure~\ref{fig:recommendation-example}.
Since the \intent{Change Analysis} intent is selected in the input panel, the system only considers the two change analysis collections: ``Summarize change in a measure between two timestamps'' (\texttt{CH1}) and ``Summarize changes in two measures over time'' (\texttt{CH2}). Next, the system enumerates possible attribute sets using the attribute mappings in Table~\ref{tab:collection-specifications} and the user's attributes of interest (\attr{Profit} and \textit{Sales}). For instance, for the \texttt{CH1} collection with the attribute mapping \{\primAttr{Q}, \primAttr{T}, \attr{4C}, \attr{G}\}, the system considers the attribute sets \{\primAttr{Profit}, \primAttr{Date}, \attr{Category}, \attr{Customer}, \attr{Product}, \attr{Region}, \attr{State}\}, \{\primAttr{Profit}, \primAttr{Date}, \attr{Category}, \attr{Product}, \attr{Region}, \attr{Segment}, \attr{State}\}, and so on.
Note that \textit{Profit} is used over \textit{Sales} as the primary attribute because it is explicitly selected by the user.

With a list of possible attribute sets ($A$) for a collection, the system then needs to select a single set ($a_{display}$) to display in its recommendations (e.g., for the \texttt{CH1} collection in Figure~\ref{fig:recommendation-example}, $a_{display} =$ \{\primAttr{Profit}, \primAttr{Date}, \attr{Category}, \attr{Region}, \attr{Segment}, \attr{ShipMode}, \attr{State}\}).
\medley~uses the following function to select the attribute set to display:

\begin{equation}
    \label{eq:interest-computation}
    a_{display} = \underset{a \in A}{max}(\frac{\sum_{i=1}^{n}v^i_{interestingness}}{n})
\end{equation}



\added{where,~$\{v^1,v^2,..., v^n\}$ are the $n$ views in a collection and $v_{interestingness}$ represents a set of normalized scores for determining the statistical interestingness of a given view ($v$).}

\added{Specifically, given a view $v$, the system \added{employs statistical metrics from prior data fact- or insight-based visualization recommendation systems (e.g.,~\cite{demiralp2017foresight,cui2019datasite,srinivasan2018augmenting,wang2019datashot,tang2017extracting}) to compute a $v_{interestingness}$ score}.
\medley~chooses metrics based on the chart type of the view, underlying data properties, and the attributes explicitly selected by the user.
The metrics include standard deviation (to determine value distributions in bar charts, maps, histograms, heatmaps), Pearson's correlation coefficient (to determine the correlation between fields in scatterplots), and smoothed z-scores (for detecting the number of peaks and drops in line charts).
To aid readability, for bar charts and heatmaps, the system also considers the cardinality of the categorical attributes, giving views with very high cardinality attributes (e.g., \attr{Customer} name) lower scores unless they are explicitly selected by the user.
The view-level metric scores are normalized to a range of $[0-1]$ across attribute sets to align scores to a common scale.
The resulting normalized $v_{interestingness}$ scores are used to determine the attribute set to display in the recommendations following Equation~\ref{eq:interest-computation}.}

By considering the intents and attributes of interest, along with these statistical metrics, \medley~generates collections that are not only relevant to the user's input (\textbf{DG2}, \textbf{DG3}) but could uncover potentially interesting data patterns (\textbf{DG4}).

\subsubsection{Collection and View Ranking}

Once \medley~generates a list of possible collections, the system ranks the collections and the views based on their relevance to the attributes of interest and the active canvas (Algorithm~\ref{algo:recommendation}, line 14).

\vspace{.5em}
\noindent\textit{Ranking collections.}
A collection's relevance is determined based on two features: an attribute match score ($C_{attrMatch}$) and a view coverage score ($C_{coverage}$).

\begin{equation}
    \label{eq:relevance-computation}
    C_{relevance} = C_{attrMatch} + C_{coverage}
\end{equation}

where $C_{attrMatch}$ is a normalized score indicating the 
\added{proportion} of intersecting attributes between a collection's primary attributes and the attributes of interest ($C_{primaryAttrs} \cap {User}_{attrs}$).
$C_{coverage}$ is a normalized score indicating the proportion of views from a collection that are already in the canvas ($C_{views} \cap {Canvas}_{views}$).

By ranking collections with a higher $C_{attrMatch}$ first, the system ensures that recommendations that focus on the attributes of interest are shown first. When the user input contains both explicit and implicit attributes, the system prioritizes explicit attributes over implicit ones (\textbf{DG2}). For example, in Figure~\ref{fig:recommendation-example}C, although both collections \texttt{CH1} and \texttt{CH2} (with the primary attributes \{\primAttr{Profit}\} and \{\primAttr{Profit}, \primAttr{Sales}\}, respectively) are equally relevant based on $C_{attrMatch}$ alone, \texttt{CH1} is ranked higher since \textit{Profit} is explicitly selected.

Combining $C_{attrMatch}$ and $C_{coverage}$ ensures that relevant collections with views containing the attributes of interest are shown first (\textbf{DG3}).
For example, in Figure~\ref{fig:interface}B, although the top four collections are comparable w.r.t. $C_{attrMatch}$, the \intent{Change Analysis} collection is shown first because two of its views (difference bar chart of \textit{Category}, map showing change in \textit{Profit}) and a widget (year picker) are already in the canvas.
Note that $C_{coverage}$ is only used when a collection contains at least one view that is not already in the canvas. 

\vspace{.5em}
\noindent\textit{Ranking views within collections.}
In addition to ranking collections, \medley~also ranks views within each collection to promote views that might be most relevant for the user to add next (\textbf{DG3}).
Specifically, for each view $v$ within a collection, the system computes a relevance score $v_{relevance}$ using the intersection between the attributes displayed in the view and the attributes of interest to the user ($v_{attrs} \cap {User}_{attrs}$).

For example, in Figure~\ref{fig:recommendation-example}C, for \texttt{CH1}, the map is shown first since there is a map already present in the canvas.
Similarly, for the \texttt{CH2}, the bar charts for \textit{Profit} and \textit{Sales} are shown first as they are attributes of interest (with \textit{Profit} being ranked higher since it is explicitly selected). \medley~moves views that are already in the canvas \added{to the end of the collection}
(e.g., the difference bar chart and map in the first collection of Figure~\ref{fig:interface}B).
\subsection{Interaction Configuration}

\medley~enables interactivity in a composed dashboard by default but also provides a lightweight GUI to edit the inferred interactions (\textbf{DG7}).
\added{To infer connections between dashboard items, the system iterates through views and widgets, treating each of them as the \emph{interaction source} and comparing them with other views and widgets on the canvas (\emph{interaction target}).
During this pairwise comparison,~\medley~uses the logic summarized in Table~\ref{tab:interaction-inference} to check if a source view/widget can be used to interactive drive a target view in the dashboard.}
Figure~\ref{fig:interactions} illustrates the three possible outcomes of the pairwise comparison.

In the first case (Figure~\ref{fig:interactions}A), when the data summary view is selected, other views are faded out to indicate that clicking on the data summary view cannot update other views (since the summary view only displays a number and cannot be used as the source to filter data in the other views).
In Figure~\ref{fig:interactions}B, when a map and bar chart are selected as the interaction [source, target] pair, a {\small{\faFilter}}~filter connector is displayed indicating that clicking on a state in the map will update categories and values in the bar chart.
Lastly, in Figure~\ref{fig:interactions}C, when the [source, target] are both maps, both {\small{\faHighlighter}}~highlight and {\small{\faFilter}}~filter connectors are shown because the views share a dimension field (\attr{State}).
When the highlight connector is active (Figure~\ref{fig:interactions}C, left), clicking on the state in the source map visually highlights the state in the target map. 
Alternatively, if the filter connector is selected (Figure~\ref{fig:interactions}C, right), all states, except the state clicked in the source view are removed from the target view.
When both {\small{\faHighlighter}}~highlight and {\small{\faFilter}}~filter interactions are possible, \medley~selects {\small{\faHighlighter}}~highlight by default since it is a less disruptive mode of interaction.

\begin{table}[t!]
\centering
\caption{\medley's interaction inference logic.}
\vspace{-.5em}
\resizebox{\linewidth}{!}{%
\begin{tabular}{@{}ll@{}}
\toprule
\textbf{Pairwise Comparison Logic}                     & \textbf{Inferred Connectors} \\ \midrule
Source view is a data summary or target view is a widget & Invalid                      \\
                                                       &                              \\
\begin{tabular}[c]{@{}l@{}}Source and target views have a shared dimension\\ OR\\ Target view encodes individual data rows (e.g., scatterplot)\end{tabular} &
  \begin{tabular}[c]{@{}l@{}}{\small{\faHighlighter}}~Highlight (default),\\ {\small{\faFilter}}~Filter\end{tabular} \\
                                                       &                              \\
All other combinations                                 & {\small{\faFilter}}~Filter                       \\ \bottomrule
\end{tabular}%
}
\label{tab:interaction-inference}
\vspace{-1em}
\end{table}

\figInteraction
\section{User Study}

We conducted a preliminary user study with \medley~to assess the utility of intent-oriented collection recommendations during dashboard composition. We had two specific goals for the study: 1) understand if and how collection recommendations aid dashboard composition and 2) gather feedback on \medley's current recommendations and features.

\figStudyResults

\subsection{Participants and Setup}
We recruited 13 participants (P1-P13, seven male, six female) through mailing lists and Slack channels containing employees from a group of visual analytics companies. The recruitment channels comprised of members in a variety of roles (e.g., manager, reporting analyst, sales) with varying levels of visualization and dashboard authoring expertise.
Participants were recruited on a first-come, first-serve basis.
When asked about their prior experience level with visualization tools like Tableau and Power BI, seven participants self-reported themselves as expert users, four participants were familiar with the general capabilities of these tools, and two participants were fairly new to visualization tools.
All participants were involved in dashboard authoring as part of their job responsibilities---the experts authored dashboards on a daily or weekly basis whereas the non-experts authored dashboards about once a month/quarter or had recently transitioned into roles that required them to start authoring dashboards. Participation in the study was voluntary, and participants were not compensated for their time.

Conforming with COVID-19 protocol, all sessions were conducted remotely via the Cisco WebEx video conferencing software~\cite{webex} and were recorded with permission from the participants.
Participants were provided with a URL to the prototype and study tasks, and completed the study on their own computers while sharing their screens for us to observe the session. All studies followed a think-aloud protocol.

\subsection{Procedure}
Sessions lasted between 40-64 minutes (mean: 55 min.) and were broken down into four parts: introduction, two tasks, and a debriefing. The study protocol and tasks were iteratively designed based on feedback from three interviewees from the formative study.

\vspace{.25em}
\noindent\textbf{Introduction} [$\sim$10min].
Participants first filled out a background questionnaire. We then played a three-minute video tutorial to familiarize participants with \medley's interface and interactions. The tutorial was intentionally designed to be brief so we could observe if participants could interpret and use the recommendations without external guidance. We also did not dive into details about the intents and objectives to prevent participant biases when performing the study tasks.

\vspace{.25em}
\noindent\textbf{Task 1: Targeted dashboard composition} [$\sim$20min].
The first task emulated a scenario of composing a dashboard based on a given criteria (e.g., a consultant authoring a dashboard based on their client's requirements). Participants were provided the Superstore dataset (described in Section~\ref{sec:usage-scenario}) and were asked to compose one or more dashboards based on two task prompts. The first task prompt, \textit{T1} was to compose a dashboard that could \textit{Summarize the differences between product segments} and the second task prompt, \textit{T2} involved composing a dashboard to \textit{Summarize the company's performance over time}.
While both \textit{T1} and \textit{T2} describe targeted dashboard goals, they were geared towards different intents (\textit{Category Analysis} and \textit{Change Analysis}, respectively) and contained different levels of attribute specificity (\textit{T1} explicitly referenced the \attr{Segment} attribute, whereas with \textit{T2}, participants were free to map ``performance'' to any one or more attributes in the dataset).
Participants were asked to refresh the window when switching between tasks or creating more than one dashboard for a single task.

\vspace{.25em}
\noindent\textbf{Task 2: Open-ended dashboard composition} [$\sim$10min].
The second task emulated an open-ended and exploratory dashboard composition scenario wherein users encounter a new dataset and iteratively formulate analytical intents as they compose the dashboard. In this task (\textit{T3}), participants were provided with two datasets: Olympic medal winners  and World Development Indicators\added{\footnote{\added{The Olympics dataset contains 15 attributes (7 \textit{Q}, 6 \textit{C}, 1 \textit{T}, and 1 \textit{G}). The World Indicators dataset contains 17 attributes (11 \textit{Q}, 4 \textit{C}, 1 \textit{T}, and 1 \textit{G}).}}}. Participants were asked to use \medley~to freely explore either (or both) datasets and compose one or more dashboards to present any notable findings.

\vspace{.25em}
\noindent\textbf{Debrief} [$\sim$10min].
This part of the study included a post-session questionnaire consisting of Likert-scale questions about the system recommendations and features.
We also included a System Usability Scale (SUS)~\cite{sus} to gauge the prototype's usability. The questionnaire was complemented with a semi-structured interview where we asked participants general questions about their overall experience, as well as targeted questions based on our observations during the session.

\subsection{Results}
\label{sec:study-results}
A total of 48 dashboards were composed during the study (mean per session: 3); some examples are shown in Figure~\ref{fig:study-results}-right.
Participants composed 2-3 dashboards during the targeted task and 1-2 dashboards during the open-ended task (seven participants only used the Olympics dataset, two only used the World Indicators dataset, and four used both).

Individual dashboards contained between 2 to 12 views and 1 to 4 widgets (mean: 5 views, 1 widget).
Nine out of the 48 participant dashboards (19\%) were directly created by adding one or two recommended collections as a whole and making minor adjustments such as removing a view or widget.
The remaining 39 dashboards (81\%) were composed using a mix of views across different collections spanning 1 to 3 intents (mean: 2) (\textbf{DG1}).
12 out of these 39 dashboards combined content from recommended collections covering the \intent{Measure Analysis} and the \intent{Category Analysis} intents. Dashboards focusing on \intent{Measure Analysis}-alone were the second-most popular category (10/39 dashboards) followed by dashboards focusing on a combination of \intent{Measure Analysis} and \intent{Change Analysis} (7/39 dashboards).


With respect to usability, on average, participants gave \medley~a SUS score of 85.96 (a score of $\geq$ 68 is considered as an indicator of good usability~\cite{sus}).
In terms of feedback on the underpinning idea of suggesting intent-based collections, participants commented favorably on both the relevance and utility of the presented recommendations.
Figure~\ref{fig:study-results}-left summarizes participant responses to a subset of the post-session questionnaire, and we discuss higher-level themes based on their responses and feedback in the subsequent section.
\section{Discussion}

\begin{figure}[t!]
        \centering
        \includegraphics[width=.75\linewidth]{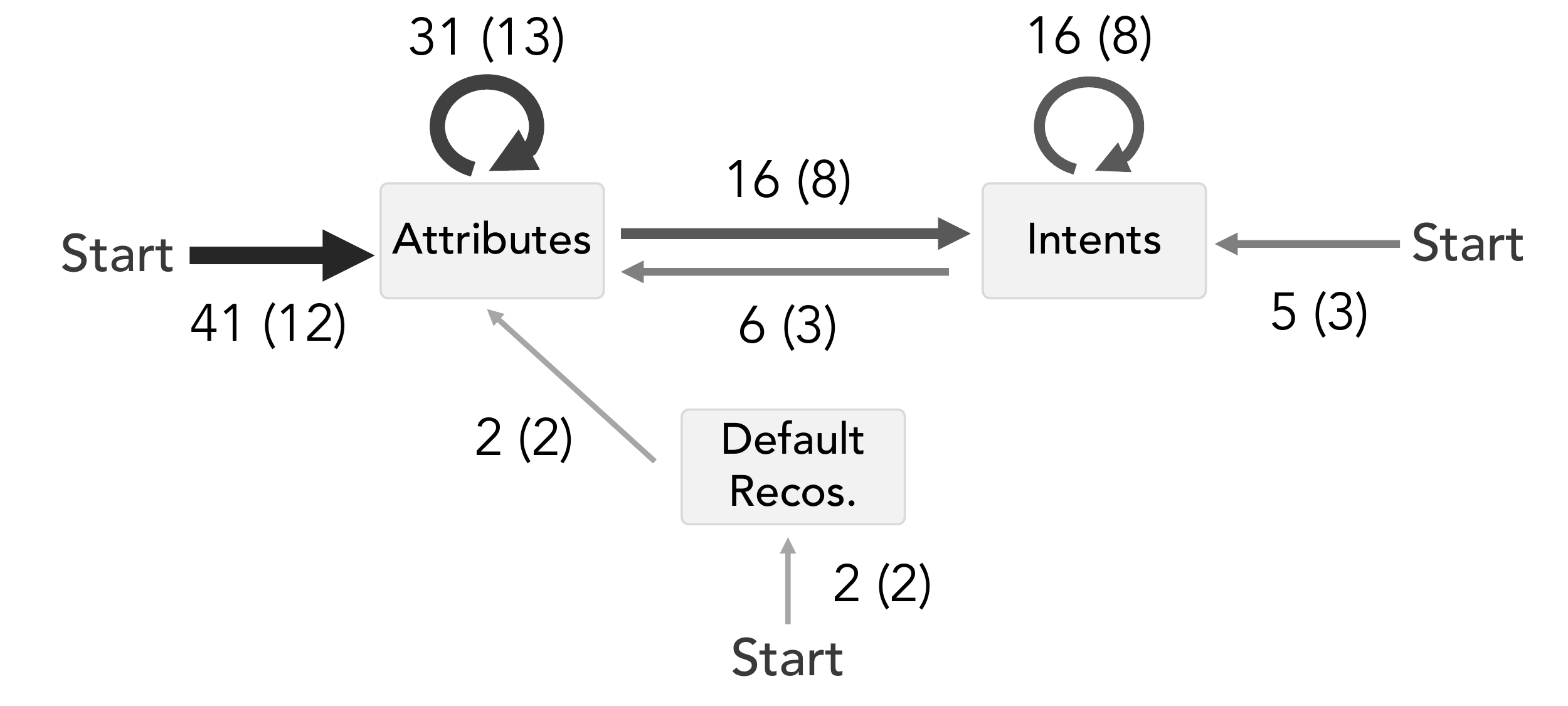}
        \vspace{-1.25em}
        \caption{Observed usage patterns during the study. Numbers on the links indicate how many of the 48 workflows---one corresponding to each of the 48 dashboards composed during the study involved a transition along with the (unique number of participants) who made that type of transition.}
        \label{fig:workflow-transitions}
        \vspace{-2.5em}
    \end{figure}

\noindent\textbf{\medley~supports flexible dashboard composition workflows.}
Participants invoked and used \medley's recommendations in a variety of ways in their dashboard composition workflows.
Figure~\ref{fig:workflow-transitions} summarizes participants' usage patterns as transitions between the actions of invoking default recommendations at the start of a session (\textbf{D}), and explicitly specifying attributes (\textbf{A}) or intents (\textbf{I}) in the input panel. In summary:

\vspace{-.5em}

\begin{itemize}[leftmargin=.1in]\itemsep-2.1pt
    \item In most cases, participants started by specifying attributes (\textit{Start}$\rightarrow$\textbf{A}, 41 instances across 12 participants). There were also five instances where participants started by specifying intents (\textit{Start}$\rightarrow$\textbf{I}) and two instances where the session started with participants using the default recommendations (\textit{Start}$\rightarrow$\textbf{D}).
    \item Upon specifying attributes, participants would typically scan the recommendations, add content to their canvas, and update the set of specified attributes to get new recommendations (\textbf{A}$\rightarrow$\textbf{A}, 31 instances).
    In some cases, instead of updating their attribute selections, participants would filter the set of recommendations by specifying one or more intents (\textbf{A}$\rightarrow$\textbf{I}, 16 instances).
    \item With intents specified, participants generally toggled through the available list of intents to focus on different types of collections (\textbf{I}$\rightarrow$\textbf{I}, 16 instances). In six cases, participants updated the attribute selections (\textbf{I}$\rightarrow$\textbf{A}) to get a new set of recommendations.
\end{itemize}

\vspace{-.25em}
\noindent{}While these patterns may be reflective of the study tasks, the breadth of patterns illustrates the potential of intent-based recommendation systems like \medley~to support a diverse set of user workflows.

\vspace{.4em}
\noindent\textbf{Recommendations facilitate rapid exploration of alternatives.}
One of the motivations for \medley~was to provide visualization recommendations to help kick-start the dashboard composition process. Participants confirmed this premise (mean agreement rating: 4.23/5), stating that the recommendations served as an easy way to scan through the views and glean dashboard alternatives.  P11 said ``\textit{My biggest point of feedback is that it definitely saved a lot of time in the upfront effort of starting a dashboard...it's more of an art to put together a dashboard so being able to explore and add an entire collection of relevant metrics and having a good first cut of a dashboard is really nice.}'' The recommendations also served as a scaffold to help users with their mental models of composing dashboards. P2 said ``\textit{Sometimes I have an idea [for a dashboard] in my head or some specific views, but sometimes I really don't, and so to see all the grouped recommendations there, it gave me more ideas of like, okay, that's a good way to look at it.}''

\vspace{.4em}
\noindent\textbf{Understanding intents and objectives needs onboarding.}
We asked participants if the concepts of intents and objectives were clear and if they found the concepts helpful as an organizing principle. While participants were in agreement about the general utility of the intents and objectives, they noted that understanding these concepts was a bit challenging at first, especially without prior onboarding. P7 commented, ``\textit{When you go into each intent and you look at the recommendations, you need to spend a few minutes to understand what is being covered, but after that, it's easy to understand and use.}'' Other participants (P4-6, P8, P12, P13) suggested adding an overlay-based onboarding tutorial to address this concern. Once participants were past the initial phase of perusing the various intents, they found it straightforward to interpret and use the recommendations (mean agreement rating: 4.08).

\vspace{.4em}
\noindent\textbf{The \medley~interface supports easy configuration of interactions.}
All 13 participants used the interaction configuration feature (Figure~\ref{fig:interactions}) to validate the connections between views, and two participants (P4, P5) used the feature to configure custom interactive behavior. All but one participant (P13) commented positively on the overlaid links and the direct manipulation-based configuration to verify and edit interactions. P13 (an expert dashboard author) had concerns about the difficulty of scaling the feature to support more complex dashboards. 
\section{Limitations and Future Work}

The user study helped identify immediate areas of improvement in \medley, including adding onboarding support for intents and improving the visibility of widgets in the recommendations. 
\added{We also identify broader limitations and topics for future work, discussed below.}

\vspace{.25em}
\noindent\added{\textbf{Enhancing recommendations.}}
\medley's current recommendations are heuristics-based, informed by a manual survey of dashboards.
We realize that more advanced recommendation systems are possible (e.g., using machine learning models on dashboard repositories) but view the design of such \added{labeled training repositories and} systems as important future research.
\added{During the study, one concern that some experts, in particular, raised about \medley's recommendations was around the lack of domain knowledge.
For instance, while working with the superstore dataset, participants wanted recommendations involving the `Profit Ratio' (a percentage value depicting $\attr{Profit}\div\attr{Sales}$).
One area for future work is to investigate how domain-specific knowledge can be incorporated into collection recommendations such that they can include derived attributes and semantically relevant visualization preferences (e.g., for an Olympics dataset, a bar chart may be a preferred default view over a map even for geographic attributes like \textit{Country}).}

\vspace{.25em}
\noindent\added{{\textbf{Expanding the evaluation.}}}
\added{We conducted a qualitative study to assess the premise of intent-driven recommendations and how \medley's design as a whole assisted dashboard composition.
As more dashboard recommendation engines are developed, future studies should focus specifically on assessing system recommendations, quantitatively evaluating their quality based on a combination of expert and crowdsourced feedback.
Another topic for quantitative evaluation is the performance of the system for different-sized datasets.
Although the template-based collection specifications and the pruning of attributes for individual views could scale to larger datasets, we have currently only tested the system with datasets containing 6-20 attributes. 
Furthermore, while the participant responses in Figure~\ref{fig:study-results}-left validate the usefulness of recommending collections during dashboard composition, additional studies are required to measure the benefits of recommending collections compared to prior single-view recommendation systems (e.g.,~\cite{wongsuphasawat2015voyager,mackinlay2007show}).
}


\vspace{-.5em}
\noindent\textbf{Combining recommendations with manual view specification.}
Prior work on visualization recommendations has demonstrated the value in combining manual view specification (MVS) and system-generated recommendations (e.g.,~\cite{wongsuphasawat2017voyager,wu2021multivision,cui2019datasite,srinivasan2018augmenting,lee2021deconstructing}).
\medley~allows users to right-click on views in the canvas and change options like the sort order, axes swapping, or data aggregation.
We intentionally chose not to support full-fledged MVS since our goal was to use \medley~as a design probe to validate the idea of leveraging intent-based collection recommendations.
Participants, however, commented that they would want the ability to specify views of interest (e.g., P10 said, ``\textit{If something pops to mind, you also want to be able to just jump to that view directly instead of scrolling through the recommendations.}"). Thus, a natural extension to our work is exploring how \medley's recommendations can be presented within the context of an MVS interface, investigating the design trade-offs, impact on recommendation generation, and potential user experience challenges of supporting the same.

\vspace{-.5em}
\noindent\removed{\textbf{Recommending dashboard layouts.}
Our goal with this initial prototype was to investigate the potential role of recommendations in helping people choose content for their dashboards.
However, besides the views themselves, the layout used to organize views is also a critical component of effective dashboard design.
Participants also noted this during our study, stating that they want suggestions not only for \textit{what} content they should add but also on \textit{how} to organize the content once they have it on the canvas.
P2, for instance, said ``\textit{A lot of times what I do now is once I have some views, I go out and look at other people’s dashboards to see how I can put things together so if the tool could also help me with that, that'd be great!}" To this end, complementing the current content-focused recommendations in~\medley, future work should explore how systems can guide expressive dashboard authoring (and not just composition) by suggesting layouts based on the semantics of the underlying dataset and/or a given set of views.}

\vspace{-.25em}
\noindent\textbf{\added{Enhancing expressivity using natural language input}.}
Five participants (P2, P3, P7, P10, and P13) suggested the possibility of using a natural language interface (NLI) as a way of easily expressing their intents and objectives. For instance, P3, said ``\textit{While this interface is powerful and easy to use, you have to know what kind of questions you need to ask in order to select those attributes and intents, and it would be really helpful to bring in something like Ask Data or Power BI Q\&A [examples of commercial NLIs] to make those selections for you.}''
Along these lines, a promising opportunity for future research is to extend NLIs for visualization (e.g.,~\cite{gao2015datatone,setlur2016eviza,narechania2020nl4dv,luo2021natural,srinivasan2021snowy}) to go beyond supporting the creation of individual views and instead support dashboard generation.
Such NLIs could enable people to use high-level utterances for stating their dashboard goals and reduce the tedium or challenges associated with explicitly selecting attributes and intents.

\vspace{-.25em}
\section{Conclusion}
Dashboards continue to be widespread and prevalent artifacts for viewing and exploring coordinated views in a variety of domains. However, authoring dashboards is often more complex and nuanced than the simple composition of pre-created views. In this work, we introduce the notion of intent-based recommendations as a scaffold for dashboard composition playing a more active part in the visual analysis process. Specifically, we describe the design and implementation of \medley, a mixed-initiative interface that supports dashboard composition by recommending dashboard collections that map to specific analytical intents. Through a usage scenario and a qualitative study with 13 participants, we demonstrate how \medley~provides interpretable recommendations that facilitate a diverse set of dashboard composition workflows.
Distilling the observations and findings from the studies, we conclude with a rich ``medley'' of future research opportunities around intelligent dashboard authoring tools, including combining manual view specification and intent-based recommendations,
and designing natural language interfaces for dashboard generation.
\acknowledgments{
We thank the anonymous reviewers and our study participants for their time and helpful feedback.
We also thank Jeff Pettiross and Megan Manetas for their detailed input and suggestions on \medley’s interface design and features.
}

\bibliographystyle{abbrv-doi}

\bibliography{references}
\end{document}